\documentstyle[12pt]{article}

\begin{document}
\title{Supersymmetric $SU(5)$ model with large supersymmetry breaking scale}
\author{N.V.Krasnikov \thanks{E-mail address: KRASNIKO@MS2.INR.AC.RU}
\\Institute for Nuclear Research\\
60-th October Anniversary Prospect 7a,\\ Moscow 117312, Russia}
\date{October,1997}
\maketitle
\begin{abstract}
Taking into account uncertainties related with the initial coupling constants 
and threshold corrections at  low and high scales we find that in 
standard supersymmetric $SU(5)$ model the supersymmetry breaking scale 
could be up to $10^8$ GeV. In the extensions of the standard supersymmetric  
 $SU(5)$ model the supersymmetry breaking scale 
could be up to $O(10^{12})$ GeV.  
In standard $SU(5)$ supersymmetric model it is possible to increase the 
GUT scale up to $5\cdot 10^{17}$ GeV provided that the masses of chiral 
superoctets and supertriplets are $m_{3,8} \sim O(10^{13})$ GeV. For $SU(5)$ 
supersymmetric model with $10^{6}$ GeV $\leq M_{SUSY} \leq$ $10^{8}$ GeV the 
Higgs boson mass is predicted to be 120 GeV $\leq m_h \leq $ 150 GeV.  
\end{abstract}
\newpage

The remarkable success of the supersymmetric $SU(5)$ model \cite{1}-\cite{12} 
is considered by many physicists as the first hint in favour of the existence 
of low energy broken supersymmetry in nature. Namely, the world 
averages with the LEP data mean that the standard nonsupersymmetric $SU(5)$ 
model \cite{13} is ruled out finally and forever (the fact that the 
standard $SU(5)$ model is in conflict with experiment was well known 
\cite{14,15} before the LEP data) but maybe the most striking and impressive 
lesson from LEP is that the supersymmetric extension of the standard 
$SU(5)$ model \cite{16}-\cite{18} predicts the Weinberg angle 
$\theta_w$ in very good agreement with experiment. The remarkable success 
of the supersymmetric $SU(5)$ model is considered by many physicists as 
the first hint in favour of the existence of low energy broken supersymmetry 
in nature. A natural question arises: is it possible to invent 
nonsupersymmetric generalizations of the standard $SU(5)$ model 
nonconfronting the experimental data or to increase the supersymmetry 
breaking scale significantly. In the $SO(10)$ model the introduction of 
the intermediate scale $M_I \sim 10^{11} GeV$ allows to obtain the Weinberg 
angle $\theta_w$ in agreement with experiment \cite{19}. In refs.\cite{20,21} 
it has been proposed to cure the problems of the standard $SU(5)$ model 
by the introduction of the additional split multiplets $5 \oplus \overline{5}$  and $10 \oplus \overline{10}$ 
in the minimal $3(\overline{5} \oplus 10)$ of the $SU(5)$ model. 
In ref.\cite{22} the extension of the standard $SU(5)$ model with 
light scalar colour octets and electroweak triplets has been proposed. 

In this paper we discuss the coupling constant unification in the standard 
supersymmetric $SU(5)$ model and its extensions. Namely, we calculate the 
masses of two key parameters of the $SU(5)$ supersymmetric  model - the mass 
of vector supermultiplet and the mass of the Higgs supertriplet. In the 
supersymmetric $SU(5)$ model both vector supermultiplet and the Higgs 
supertriplet are responsible for the proton decay.   Taking into account the   
uncertainties associated with the initial gauge coupling constants 
and threshold corrections we conclude 
that in the standard supersymmetric $SU(5)$ model the scale of the 
supersymmetry breaking could be up to $10^8$ GeV. We find  that 
in the extensions of the standard $SU(5)$ supersymmetric model it is 
possible to increase the supersymmetry breaking scale up to $10^{12}$ GeV. 
In standard  $SU(5)$ supersymmetric model it is possible to increase the 
GUT scale up to $5 \cdot 10^{17}$ GeV provided that the masses of chiral 
superoctets and supertriplets are $m_{3,8} \sim O(10^{13})$ GeV. For the 
supersymmetric $SU(5)$ model with the supersymmetry breaking scale  
$10^{6}$ GeV $ \leq M_{SUSY} \leq 10^8 $ GeV the Higgs boson mass is predicted 
 to be 130 GeV $ \leq m_h \leq $ 150 GeV. Some of the results presented 
here are contained in our previous papers \cite{23}.

The standard supersymmetric $SU(5)$ model \cite{16}-\cite{18} contains three 
light supermatter generations and two light superhiggs doublets. 
A minimal choice of massive supermultiplets at the high scale is 
$(\overline{3},2,\frac{5}{2}) \oplus c.c.$ massive vector supermultiplet 
with the mass $M_v$, massive chiral supermultiplets $(8,1,0), (1,3,0), (1,1,0)$ 
with the masses $m_8, m_3, m_1$ (embeded in a 24 supermultiplet of $SU(5)$) 
and a $(3,1,-\frac{1}{3}) \oplus (-3,1,\frac{1}{3})$ complex Higgs 
supermultiplet with a mass $M_3$ embeded in $5 \oplus \overline{5}$ of $SU(5)$. 
In low energy spectrum we have squark and slepton multiplets 
$ (\tilde{u},\tilde{d})_{L}, \tilde{u}^c_L, \tilde{d}^c_L, 
(\tilde{\nu},\tilde{e})_L, \tilde{e}^c_L $ plus the corresponding 
squarks and sleptons of the second and third supergenerations. 
Besides in the low energy spectrum we have $SU(3)$ octet of gluino 
with a mass $m_{\tilde{g}}$, triplet of $SU(2)$ gaugino with a mass 
$m_{\tilde{w}}$ and the photino with a mass $m_{\tilde{\gamma}}$. 
For the energies between $M_z$ and $M_{GUT}$ we have effective 
$SU(3) \otimes SU(2) \otimes U(1)$ gauge theory. 
In one loop approximation the corresponding solutions of the renormalization 
group equations are well known \cite{18}. In our paper instead of the 
prediction of $\sin^{2}(\theta_w)$ following refs.\cite{6}-\cite{23,24}  
we consider the 
following one loop relations between the effective gauge coupling constants,
the mass of the vector massive supermultiplet $M_v$ and the mass of the 
superhiggs triplet $M_3$:
\begin{equation}
A \equiv  2(\frac{1}{\alpha_{1}(m_{t})} - \frac{1}{\alpha_{3}(m_t)}) + 
3(\frac{1}{\alpha_{1}(m_t)} - \frac{1}{\alpha_{2}(m_t)}) = \Delta_{A} ,
\end{equation}
\begin{equation}
B \equiv  2(\frac{1}{\alpha_{1}(m_t)} - \frac{1}{\alpha{3}(m_t)}) - 
3(\frac{1}{\alpha_{1}(m_t)} - \frac{1}{\alpha_{2}(m_t)}) = \Delta_{B} ,
\end{equation}
where
\begin{equation}
\Delta_{A} = (\frac{1}{2\pi})(\delta_{1A} + \delta_{2A} + \delta_{3A}) ,
\end{equation}
\begin{equation}
\Delta_{B} = (\frac{1}{2\pi})(\delta_{1B} + \delta_{2B} + \delta_{3B}) ,
\end{equation}
\begin{equation}
\delta_{1A} = 44ln(\frac{M_v}{m_t}) - 4ln(\frac{M_v}{m_{\tilde{g}}}) -
4ln(\frac{M_v}{m_{\tilde{w}}}) ,
\end{equation}
\begin{equation}
\delta_{2A} = -12(ln(\frac{M_v}{m_8}) + ln(\frac{M_v}{m_3})) ,
\end{equation}
\begin{equation}
\delta_{3A} = 6ln(m_{(\tilde{u},\tilde{d})_L}) - 3ln(m_{\tilde{u}^c_L}) - 
3ln(m_{\tilde{e}^c_L}) ,
\end{equation}
\begin{equation}
\delta_{1B} = 0.4ln(\frac{M_3}{m_h}) + 0.4ln(\frac{M_3}{m_H}) + 
1.6ln(\frac{M_3}{m_{sh}}) ,
\end{equation}
\begin{equation}
\delta_{2B} = 4ln(\frac{m_{\tilde{g}}}{m_{\tilde{w}}}) + 
6ln(\frac{m_8}{m_3}) ,
\end{equation}
\begin{equation}
5\delta_{3B} = -12ln(m_{(\tilde{u},\tilde{d})_L}) + 9ln(m_{\tilde{u}^c_L}) + 
6ln(m_{\tilde{d}^c_L}) - 
6ln(m_{(\tilde{\nu} ,\tilde{e})_L}) + 3ln(m_{\tilde{e}^c_L}) 
\end{equation}
Here $m_h$, $m_H$ and $m_{sh}$ are the masses of the first light Higgs 
isodoublet, the second Higgs isodoublet and the isodoublet of superhiggses.
The relations (1-10) are very convenient since they allow to determine 
separately two key parameters of the high energy spectrum of $SU(5)$ model, 
the mass of the vector supermultiplet $M_v$ and the mass of the chiral 
supertriplet $M_3$. Both the vector supermultiplet and the chiral supertriplet 
are responsible for the proton decay in supersymmetric $SU(5)$ model \cite{18}.
In the standard nonsupersymmetric $SU(5)$ model the proton lifetime 
due to the massive vector exchange is determined by the formula \cite{25}
\begin{equation}
\Gamma(p \rightarrow e^{+} \pi^{o})^{-1} = 4 \cdot 10^{29 \pm 0.7}
(\frac{M_v}{2 \cdot10^{14} Gev})^{4} yr
\end{equation}
In the supersymmetric $SU(5)$ model the GUT coupling constant is 
$\alpha_{GUT} \approx \frac{1}{25}$ compared to 
$\alpha_{GUT} \approx \frac{1}{41}$ in standard $SU(5)$ model, so we 
have to multiply the expression (11) by factor $(\frac{25}{41})^2$. 
From the current experimental limit \cite{24}
$\Gamma(p \rightarrow e^{+} \pi^{o})^{-1} \geq 5.5 \cdot 10^{32} yr $ 
we conclude that $M_v \geq 1 \cdot 10^{15} Gev$. The corresponding 
experimental bound on the mass of the superhiggs triplet $M_3$ depends 
on the masses of gaugino and squarks \cite{25,26}. In our calculations 
we use the following values for the initial coupling constants \cite{26},
\cite{29}-\cite{32}:
\begin{equation}
\alpha_{3}(M_z) = 0.118 \pm 0.003 ,
\end{equation}
\begin{equation}
\sin^{2}_{\overline{MS}}(\theta_w)(M_z) = 0.2320 \pm 0.0005 ,
\end{equation}
\begin{equation}
(\alpha_{em,\overline{MS}}(M_z))^{-1} = 127.79 \pm 0.13 
\end{equation}
For the top quark mass $m_t = 174 GeV$ after the solution of the corresponding 
renormalization group equations in the region $ M_z \leq E \leq m_t$ we find 
that in the $\bar{MS}$-scheme
\begin{equation}
A_{\bar{MS}} = 183.96 \pm 0.47 ,
\end{equation}
\begin{equation}
B_{\bar{MS}} = 13.02 \pm 0.45 . 
\end{equation}
Here the  errors in formulae (15), (16) are determined mainly by the error 
in the determination of the strong coupling constant $\alpha_{s}(M_{Z})$. 
Since we study the $SU(5)$ supersymmetric model the more appropriate is to 
use the $\bar{DR}$-scheme. The relation between the coupling constants in the 
$\bar{MS}$- and $\bar{DR}$-schemes has the form \cite{34} 
\begin{equation}
\frac{1}{\alpha_{i_{\bar{MS}}}} = \frac{1}{\alpha_{i_{\bar{DR}}}} + 
\frac{C_{2}(G)}{12\pi}, 
\end{equation}
where $C_{2}(G)$ is the quadratic Casimir operator for the adjoint 
representation. In the $\bar{DR}$-scheme we find that 
\begin{equation}
A_{\bar{DR}} = A_{\bar{MS}} + \frac{1}{\pi} = 184.28 \pm 0.47 ,
\end{equation}
\begin{equation}
B_{\bar{DR}} = B_{\bar{MS}} = 13.02 \pm 0.45 .
\end{equation}
 
Using one loop formulae (1)-(10) in the neglection of the contributions due 
to sparticle mass differences and high scale threshold corrections 
($\delta_{2A} = \delta_{2B} = \delta_{3A} = \delta_{3B} = 0 $) we find 
\begin{equation}
M_{v} = 1.8 (\frac{175 GeV}{M_{SUSY}})^{\frac{2}{9}} \cdot 
10^{16 \pm 0.04} GeV ,
\end{equation}
\begin{equation}
M_{3} = 1.1 \frac{M_{h,eff}}{175 GeV} \cdot 10^{17 \pm 0.5} GeV ,
\end{equation}  
where $M_{SUSY} \equiv (m_{\tilde{g}} m_{\tilde{w}})^{\frac{1}{2}}$  and 
$M_{h,eff} \equiv (m_h m_H)^{\frac{1}{6}} \times m_{sh}^{\frac{2}{3}}$.                            

An account of two loop corrections leads to 
the appearance of the additional factors
\begin{equation}
2 \pi  \delta_{4A,4B} = 2(\theta_1 - \theta_3) \pm 3(\theta_1 - 
\theta_3) ,
\end{equation}
\begin{equation}
\theta_{i} = \frac{1}{4\pi}\sum_{j=1}^{3} 
\frac{b_{ij}}{b_{j}}ln[\frac{\alpha_{j}(M_v)}
{\alpha_{j}(m_t)}]
\end{equation}
Here $b_{ij}$ are the two loop $\beta$-functions  coefficients.  
An account of two loop corrections (22) for $M_{SUSY} \sim 500 $ GeV 
leads to the increase of $M_{v}$ by factor 1.2. and to the decrease of 
$M_3$ by factor 56. An account of two-loop corrections (22) due to 
nonzero top quark Yukawa coupling constant as it has been found in ref.
\cite{11} leads to the additional negative corrections to the one loop 
beta function coefficients
\begin{equation}
b_i \rightarrow b_i - b_{i;top}\frac{h^2_t}{16\pi^2} ,
\end{equation}
where $b_{i,top} = \frac{26}{5}, 6, 4 $ for $i = 1,2,3$. We have found that 
an account of two loop Yukawa corrections practically does not change 
the value of $M_v$ and leads to the small increase of $M_3$ by factor 
1.5(1.2) for $h_t(m_t)=1(h_t(m_t)=0.8)$. In the assumption that all gaugino 
masses coincide at GUT scale we find standard relation
\begin{equation}
m_{\tilde{g}} = \frac{\alpha_3(M_{SUSY})}{\alpha_2(M_{SUSY})}m_{\tilde{w}} = 
(2-2.5)m_{\tilde{w}}
\end{equation}
between gluino and wino masses that leads to the decrease of $M_3$ by 
factor 3.2-4.6. We have found that the values of $\delta_{3B}(\delta_{3A})$ 
for realistic spectrum are between 0 and 0.4 (0 and 5.5) that leads to the 
maximal decrease of $M_3(M_v)$ by factor 1.5(1.4). Taking into account these 
corrections we find that 
\begin{equation}
M_v = 2.0 \cdot (1-0.67) \cdot (\frac{175 GeV}{M_{SUSY}})^{\frac{2}{9}} \cdot 
10^{16 \pm 0.04 GeV},
\end{equation}
\begin{equation}
M_3 = 0.80 \cdot (1-0.43) \cdot \frac{M_{h,eff}}{175 GeV} 10^{15 \pm 0.5}
\end{equation}
It should be noted that estimates (25), (26) are obtained in the assumption 
$M_v=m_3=m_8$. From the lower bound $10^{15}GeV$ on the value of the mass of 
the vector bosons responsible for the baryon number nonconservation we find 
an upper bound on the supersymmetry breaking scale $M_{SUSY} \leq 2 \cdot 
10^{18} GeV$. Let us consider now Eqs. (2), (8), (10), (26). For the lightest 
Higgs boson mass $m_h=100 GeV$ in the assumption that $m_{H} = m_{sh} = 
M_{SUSY}$ and $M_3 \leq 3 M_v$ \cite{5} \footnote{The inequality 
$M_3 \leq 3 M_v$ 
comes from the requirement of the absence of Landau Pole singularities for 
effective charges for energies up to Planck scale.} we find that 
\begin{equation}
M_{SUSY} \leq 50 TeV
\end{equation}
Let us stress that the inequality (27) is in fact an upper bound for the 
second superhiggs doublet masses. For the standard case when at GUT scale 
we have universal soft supersymmetry breaking mass terms for squarks, 
sleptons  and Higgses after solution of the corresponding renormalization 
group equations for soft supersymmetry breaking parameters  
squarks, sleptons and higgses masses coincide up to factor 3. In the general 
case the masses of Higgses, squarks and sleptons are arbitrary parameters. 
The masses of shiggses are determined by the mu-parameter which is also a 
free parameter of the model. In the assumption that the shiggs mass is 
$m_{sh} \leq O(1) TeV$ and $M_H = M_{SUSY}$ we have found that the inequality 
$M_3 \leq 3M_v$ is satisfied for the supersymmetry breraking scale up to 
$10^8 GeV$.  

The proton lifetime due to the exchange of the Higgs supertriplet is 
determined by  the mode $p \rightarrow \bar{\nu} K^{+}$ . 
 From the nonobservation of 
$p \rightarrow \bar{\nu} K^{+}$ decay Arnowitt and Nath \cite{5} derived 
an upper limit on the parameter C which can be written in the form
\begin{equation}
C \leq 335 \cdot \frac{M_3}{6 \cdot 10^{16} GeV},
\end{equation}
\begin{equation}
C = \frac{-2\alpha_{2}}{\alpha_{3}\sin(2\beta)} \frac{m_{\tilde{g}}}
{m^2_{\bar{q}} \cdot 10^{-6} GeV^{-1}} .
\end{equation}
From Eqs. (26), (28), (29) we find that 
\begin{equation}
\frac{m_{\tilde{g}}}{m^2_{\tilde{q}}M_{h,eff}} \leq 84 \cdot 10^{-9} GeV^{-2} .
\end{equation}
From the inequality (30) and from the experimental bound \cite{33} 
$m_{\tilde{g}} \geq 170 GeV$ on the gluino mass in the assumption that 
$m_{\tilde{q}} = m_{H} = m_{sh}$ we find bound on the squark mass 
\begin{equation}
m_{\tilde{q}} \geq 1460 GeV .
\end{equation}
As it has been mentioned previously the main uncertainty in the numerical 
determination of the B parameter results from the uncertainty in the 
determination of $\alpha_{s}(M_Z)$. For $\alpha_{s}(M_Z) = 0.121$ we have 
bound (31). For $\alpha_{s}(M_{Z}) = 0.118(0.115)$ we find that 
$ m_{\tilde{q}} 
\geq $ 2200 GeV(3300 GeV). It should be noted that up to now we assumed that 
at the GUT scale all gaugino masses coincide. If we refuse from this 
requirement it is possible to increase the supertriplet mass $M_3$ since 
$M_3$ is proportional to $ (\frac{m_{\tilde{w}}}{m_{\tilde{g}}})^{\frac{5}{3}}$. 
For instance, for $m_{\tilde{g}} = m_{\tilde{w}}$ we find that $M_3$ could 
be up to $5.4 \cdot 10^{16} GeV$ and as a consequence we have more weak bound 
$m_{\tilde{q}} \geq 900 GeV$ for the squark mass. Besides for 
$m_8 \neq m_3$ we have additional factor $(\frac{m_3}{m_8})^{2.5}$ in front of 
the expression for the determination of $M_3$ that allows to increase the 
value of $M_3 $. 

Another way to satisfy the equation (2) is the introduction of the additional 
Higgs superdoublets. Namely, let us add two additional superhiggs 5-plets. 
Assume that after $SU(5)$ gauge symmetry breaking the light Higgs isodoublet 
has a mass $O(M_Z)$, other  Higgs isodoublets  and superhiggses have 
masses $O(M_{SUSY})$. For such case the solution of the modified equation 
(2) allows to determine the Higgs supertriplet mass. For $m_t =175 GeV$, 
$m_8 = m_3 = m_v$ and for $\alpha_s(M_Z) = 0.118$ we have found that 
for $ M_{SUSY} = 10^{7} GeV(10^{6} GeV)$ the vector boson masses and the 
Higgs supertriplet masses are

$M_v = 1.3 \cdot 10^{15} GeV (2.2 \cdot 10^{15} GeV)$,

$M_3 = 1.4 \cdot 10^{13} GeV(1.7 \cdot 10^{12} GeV)$.

For such SUSY breaking scale  $M_{SUSY}$  the $d = 5$ operators leading 
to fast proton decay are not dangerous provided that $m_{\tilde{q}} 
\gg  m_{\tilde{g}}$.

It is instructive to consider the supersymmetric $SU(5)$ model with 
relatively light colour octet and triplets. For instance, consider the 
superpotential
\begin{equation}
W = \lambda \sigma(x)[Tr(\Phi^{2}(x)) -c^2] ,
\end{equation}
where $\sigma(x)$ is the $SU(5)$ singlet chiral superfield and $\Phi(x)$ is 
chiral 24-plet in the adjoint representation. For the superpotential (33) 
the colour octet and electroweak triplet chiral superfields remain 
massless after $SU(5)$ gauge symmetry breaking and they acquire the 
masses $O(M_{SUSY})$ after the supersymmetry breaking. So in this scenario 
we have additional relatively light fields. Lower bound on the mass 
of the vector bosons leads to the upper bound on the supersummetry 
breaking scale 
$M_{SUSY} \leq 10^{12} GeV $.  In order to satisfy 
the second equation for the mass of the Higgs triplets 
let us introduce in the model four  additional superhiggs 5-plets. 
If we assume 
that after $SU(5)$ gauge symmetry breaking the corresponding Higgs triplets 
acquire mass $O(M_v)$ , light  Higgs isodoublet has a 
mass $O(M_z)$ , other  Higgs isodoublets and superhiggses have masses 
$O(M_{SUSY})$ then we can satisfy the equation (2) for 
$M_{SUSY} \sim 10^{12} GeV$.

In standard supersymmetric $SU(5)$ model the superpotential containing 
the selfinteraction of the chiral 24-supermultiplet has the form
\begin{equation}
W(\Phi(x)) = \lambda [Tr(\Phi(x)^{3}) + MTr(\Phi(x)^{2})]
\end{equation}
The vacuum solution 
\begin{equation}
\Phi(x) = \frac{4M}{3}Diag(1,1,1,-\frac{3}{2}, -\frac{3}{2})
\end{equation}
leads to the $SU(5) \rightarrow SU(3) \otimes SU(2) \otimes U(1)$ gauge 
symmetry breaking. After the gauge symmetry breaking the octets and triplets 
acquire masses $m_8 = m_3 = 10 \lambda M$. So in general the superoctet and 
the supertriplet masses do not coincide with the vector boson mass $M_v$ 
and in fact they are free parameters of the model. It is possible to 
have grand unified scale $M_v = 5\cdot 10^{17} GeV$ and $M_{SUSY} \leq 
1 TeV$  \cite{23} provided that the octets and the triplets 
are lighter than the 
vector supermultiplet by factor 15000 which is welcomed from the superstring 
point of view \cite{35}. Playing with $m_8 = m_3 \neq m_v$ it is also 
possible to increase the value of SUSY breaking scale. For instance, 
for $m_8 = m_3 = 10^{-2}m_v$ the SUSY breaking scale is increased by 
factor $10^3$ compared to the case $m_v = m_8 = m_3$. For $M_{SUSY} = 
10^{9} GeV$ it is possible to satisfy the equation (2) by the inroduction 
of 2 additional relatively light superdoublets with the masses $O(M_{SUSY})$. 
In this case the value of the Higgs supertriplet mass is predicted to be 
$O(10^{15})GeV$ and $d=5$ operators are not dangerous. 

In $SU(5)$ supersymmetric model with relatively big SUSY breaking scale it is 
possible to predict the mass range interval for the lightest Higgs boson. 
Really, at scale $M \geq O(M_{SUSY})$ we have minimal standard supersymmetric 
model(MSSM) and at lower scales standard Weinberg-Salam model works. As a 
consequence the effective Higgs 
selfcoupling constant for the Weinberg-Salam model at the supersymmetry 
breaking scale has to obey the inequality \cite{37}
\begin{equation}
0 \leq \bar{\lambda}(M_{SUSY}) \leq \frac{ \bar{g}^2_1(M_{SUSY}) + 
\bar{g}^2_2(M_{SUSY})}{4}
\end{equation}
So the assumption that standard Weinberg-Salam model originates from its 
supersymmetric extension with the supersymmetry breaking scale $M_{SUSY}$ 
allows to obtain non-trivial information about the low energy effective 
Higgs self-coupling constant in the effective potential 
$V = - M^2H^{+}H + \frac{\lambda}{2}(H^{+}H)^2$ and hence to obtain 
nontrivial information on the Higgs boson mass. The corresponding calculations 
 have been done in ref.\cite{37}. Numerically, for $m_t = 175 GeV$, 
$\alpha_{s}(M_Z) =0.118$, $10^{6} GeV \leq M_{SUSY} \leq 10^{8} GeV$ 
the Higgs boson mass is predicted to be $120 GeV \leq m_h \leq 150 GeV$. 
The dependence of the Higgs boson mass on the supersymmetry breaking scale 
$M_{SUSY}$ is rather weak for $M_{SUSY} \geq 10^{6} GeV$.

In conclusion let us formulate our main results. In standard $SU(5)$ 
supersymmetric model we have calculated the masses of vector supermultiplet 
and of the Higgs supertriplet.  We have found that 
in standard supersymmetric $SU(5)$ model with colour octet 
and triplet masses $O(M_v)$ and  equal gaugino masses at GUT scale 
the nonobservation of the proton decay 
leads to the upper bound $M_{SUSY} \leq  10^{8} Gev$ on the   
supersymmetry breaking scale. For the case when octets and triplets have 
masses $O(M_{SUSY})$ it is possible to increase the supersymmetry 
breaking scale up to $O(10^{12}) GeV$, however, in this case in order to 
satisfy the  Eq. (2) for the superhiggs triplet mass we have to introduce 
additional relatively light superhiggs doublets. We have demonstrated also 
that in standard $SU(5)$ model it is possible to have a GUT scale $M_v \sim 
5 \cdot 10^{17} GeV$ and the supersymmetry breaking scale $M_{SUSY} \leq 
1 TeV$ provided that the octets and the triplets are lighter than the vector 
supermultiplet by factor $O(15000)$. It should be noted that the obtained 
bound on the supersymmetry breaking scale $M_{SUSY}$ depends rather 
strongly on the details of the high energy spectrum and on the splitting 
between gaugino masses. For $10^6 GeV \leq M_{SUSY} \leq 10^{8} GeV$ the 
Higgs boson mass is predicted to be $130 GeV \leq m_h \leq 150 GeV$. 
For $M_{SUSY} \geq O(1) TeV$ we have a fine tuning problem for the 
electroweak symmetry breaking scale.

I am indebted to the collaborators of the INR theoretical department 
for discussions and critical comments. The research described in this 
publication was made possible in part by Award No RPI-187 of the U.S. 
Civilian Research and Development Foundation for the Independent States 
of the Former Soviet Union(CRDF).


\newpage
\begin{thebibliography}{99}
\bibitem{1} U.Amaldi, W.de Boer and H.Furstenau, Phys.Lett.{\bf B260}(1991)
447; \\
C.Guinti, C.W.Kim and U.W.Lee, Mod.Phys.Lett.{\bf A6}(1991)1745.
\bibitem{2} J.Ellis, S.Kelley and D.V.Nanopoulos, Phys.Lett.{\bf B260}(1991)
131.
\bibitem{3} P.Langacker and M.Luo, Phys.Rev.{\bf D44}(1991)817.
\bibitem{4} G.G.Ross and R.G.Roberts, Nucl.Phys.{\bf B377}(1992)571.
\bibitem{5} R.Arnowith and P.Nath, Phys.Rev.Lett.{\bf 69}(1992)725.
\bibitem{6} J.Hisano, H.Murayama and T.Yanagida, Phys.Rev.Lett.{\bf 69}(1992)
1014.
\bibitem{7} S.Keley et al., Phys.Lett.{\bf B273}(1991)423.
\bibitem{8} R.Barbieri and L.J.Hall, Phys.Rev.Lett.{\bf 68}(1992)752.
\bibitem{9} J.Ellis, S.Keley and D.V.Nanopoulos, Nucl.Phys.{\bf B373}(1992)55.
\bibitem{10} K.Hagiwara and Y.Yamada, Phys.Rev.Lett.{\bf 70}(1993)709.
\bibitem{11} P.Langacker and N.Polonsky, Phys.Rev.{\bf D47}(1993)4028.
\bibitem{12} V.Barger, M.S.Berger and P.Ohmann, Phys.Rev.{\bf D47}(1993)1093.
\bibitem{13} H.Georgi and S.L.Glashow, Phys.Rev.Lett. {\bf 32}(1974)438;
H.Georgi, H.R.Quinn and S.Weinberg, Phys.Rev.Lett.{\bf 33}(1974)451.
\bibitem{14} U.Amaldi et al., Phys.Rev.{\bf D36}(1987)1385.
\bibitem{15} G.Costa et al., Nucl.Phys.{\bf B297}(1988)244.
\bibitem{16} S.Dimopoulos, S.Raby and F.Wilczek, Phys.Rev.{\bf D24}(1981)1681.
\bibitem{17} M.B.Einhorn and D.R.T.Jones, Nucl.Phys.{\bf B196}(1982)475.
\bibitem{18} For reviews, see: H.P.Nilles, Phys.Rep.{\bf 110}(1984)1;
G.G.Ross, Grand Unified Theories (Benjamin, New York 1984); 
R.N.Mohapatra, Unification and Supersymmetry (Springer, New York 1992).
\bibitem{19}  J.C.Pati, Abdus Salam and J.Strathdee, Nucl.Phys.{\bf B185}
(1981)445.
\bibitem{20} P.H.Frampton and S.L.Glashow, Phys.Lett.{\bf B131}(1983)340.
\bibitem{21} U.Amaldi et al., Phys.Lett.{\bf B281}(1992)374.
\bibitem{22} N.V.Krasnikov, Phys.Lett.{\bf B306}(1993)283.
\bibitem{23} N.V.Krasnikov, Pis'ma ZhETP{\bf 61}(1995)236; \\
N.V.Krasnikov, Phys.Lett.{\bf B392}(1997)365.
\bibitem{24} J.Hisano, T.Moroi, K.Tobe and T.Yanagida, Mod.Phys.Lett.
{\bf A10}(1995)2267.
\bibitem{25} See for instance: W.Marciano, in: Proc. 8th Grand Unification 
Workshop (Syracuse, N.Y., 1987), ed. K.G.Wali, (World Scientific Singapore, 
1988) p.185.
\bibitem{26} Particle Data Group, Review of particle properties, 
Phys.Rev.{\bf D50}(1994).
\bibitem{27} N.Sakai and T.Yanagida, Nucl.Phys.{\bf B197}(1982)533.
\bibitem{28} S.Weinberg, Phys.Rev.{\bf D46}(1982)287.
\bibitem{29} N.Schmelling,  in: Proc. 28th Int. Conf. on High Energy Physics,\\
ed. Z.Ajduk and A.K.Wroblewski(World Scientific, Singapore, 1997) p.91.
\bibitem{30} P.Langacker and N.Polonsky, Phys.Rev.{\bf D52}(1995)3081.
\bibitem{31} P.Chanowski, Z.Pluciennik and S.Pokorski, Nucl.Phys.{\bf B439}
(995)23.
\bibitem{32} N.V.Krasnikov, Mod.Phys.Lett.{\bf A9}(1994)2825.
\bibitem{33} P.Mattig, in: Proc. 28th Int. Conf. on High Energy Physics, \\
ed. Z.Ajduk and A.K.Wroblewski(World Scientific, Singapore 1997) p.309.
\bibitem{34} I.Antoniadis, C.Kounnas and K.Tamvakis, Phys.Lett.{\bf B119}
(1982)377.
\bibitem{35} V.Kaplunovski, Nucl.Phys.{\bf B307}(1988)145.
\bibitem{36} N.V.Krasnikov, Mod.Phys.Lett.{\bf A10}(1995)2675.
\bibitem{37} N.V.Krasnikov, G.Kreyerhoff and R.Rodenberg, Mod.Phys.Lett.
{\bf A9}(1994)3663.
    
\end{thebibliography}
\end{document}